\documentclass[lettersize,journal]{IEEEtran}
\usepackage{amsmath,amsfonts}
\usepackage{algorithmic}
\usepackage{algorithm}
\usepackage{array}
\usepackage[caption=false,font=normalsize,labelfont=sf,textfont=sf]{subfig}
\usepackage{textcomp}
\usepackage{stfloats}
\usepackage{url}
\usepackage{verbatim}
\usepackage{graphicx}
\usepackage{cite}
\usepackage{soul}
\usepackage{color, xcolor} 

\begin{document}

\title{Multi-Modal Intelligent Channel Modeling Framework for 6G-Enabled Networked Intelligent Systems}

\author{Lu~Bai,~\IEEEmembership{Senior~Member,~IEEE}, Zengrui~Han,~\IEEEmembership{Student~Member,~IEEE}, Xuesong~Cai,~\IEEEmembership{Senior~Member,~IEEE}, and Xiang~Cheng,~\IEEEmembership{Fellow,~IEEE}

\thanks{L.~Bai is with the Joint SDU-NTU Centre for Artificial Intelligence Research (C-FAIR), Shandong University, Jinan, 250101, P. R. China (e-mail: lubai@sdu.edu.cn).}
\thanks{Z. Han, X. Cai, and X.~Cheng are with the State Key Laboratory of Photonics and Communications, School of Electronics, Peking University, Beijing, 100871, P. R. China (email: zengruihan@stu.pku.edu.cn, xuesong.cai@pku.edu.cn, xiangcheng@pku.edu.cn).}
}



\maketitle

\begin{abstract}
The design and technology development of 6G-enabled networked intelligent systems needs an accurate real-time channel model as the cornerstone. However, with the new requirements of 6G-enabled networked intelligent systems, the conventional channel modeling methods face many limitations. Fortunately, the multi-modal sensors equipped on the intelligent agents bring timely opportunities, i.e., the intelligent integration and mutually beneficial mechanism between communications and multi-modal sensing could be investigated based on the artificial intelligence (AI) technologies. In this case, the mapping relationship between physical environment and electromagnetic channel could be explored via Synesthesia of Machines (SoM). This article presents a novel multi-modal intelligent channel modeling (MMICM) framework for 6G-enabled networked intelligent systems, which establishes a nonlinear model between multi-modal sensing and channel characteristics, including large-scale and small-scale channel characteristics. The architecture and features of proposed intelligent modeling framework are expounded and the key technologies involved are also analyzed. Finally, the system-engaged applications and potential research directions of MMICM framework are outlined. 
\end{abstract}

\begin{IEEEkeywords}
6G-enabled networked intelligent systems, multi-modal intelligent channel modeling, multi-modal sensing, synesthesia of machine. 
\end{IEEEkeywords}

\section{Introduction}
With the rapid development of artificial intelligence technology, the research of intelligent unmanned systems has attracted more and more attention while it supports many applications that promote the development of the national economy, e.g., intelligent transportation systems, unmanned storage and logistics systems, etc. 
In consideration of ability limitation of single intelligent agent, the research of intelligent networks for intelligent unmanned systems is also significant, which furnishes an efficient and reliable communication network among intelligent agents to carry out large volume unmanned tasks \cite{NIS}. 
As we all know, channel modeling is the cornerstone of design and technology development of wireless communications \cite{6}. Therefore, an accurate and easy-to-use channel model is essential for 6G-enabled networked intelligent systems.

However, the research on channel modeling for 6G-enabled networked intelligent systems is encountering new and severe demand challenges, such as precise prediction capability, extension capabilities at diverse scenarios and frequency bands, and system participation capability, which existing channel modeling methods cannot satisfied. 
Deterministic channel modeling illustrates the channel under site-specific scenarios in a deterministic manner, where the electromagnetic wave propagation mechanism can be effectively simulated. However, the detailed and site-specific representation results in significant computational complexity in deterministic channel models.
To mitigate modeling complexity, stochastic channel modeling characterizes radio propagation statistically, leveraging probability density functions (PDFs) of channel parameters derived from measurement data. Being statistically based, this channel modeling method maintains stochastic characteristics while offering constrained precision. 
Meanwhile, as communication and intelligence become integrated, future systems are expected to be AI-native. In this context, the scale and fidelity of available datasets are critical, as they fundamentally determine the upper performance limit of these AI-native communication systems. However, conventional approaches struggle to meet this demand.
With the advancement of artificial intelligence (AI) technologies, the application of AI technologies has been incorporated into channel modeling studies, injecting new vitality into channel modeling methodologies. Ref. \cite{survey-AI} presents a comprehensive survey of AI-based channel modeling approaches developed in recent years. 
Nevertheless, constrained by the exclusive use of radio frequency (RF) data and panoramic maps
for AI-based channel modeling, related research remains inadequate in a detailed perception of the physical environment. Consequently, they fail to meet the stringent requirements of high dynamic characteristics and ultra-reliable low-latency communications in 6G-enabled networked intelligent systems. Fortunately, the deployment of multi-modal sensors on intelligent agents offers a transformative opportunity to investigate AI-driven mutually beneficial mechanism between communication and multi-modal sensing. This framework enables the investigation of the mapping relationship between the physical environment and electromagnetic channel characteristics via the paradigm of Synesthesia of Machines (SoM).

Therefore, this paper investigates a novel multi-modal intelligent channel modeling (MMICM) framework for 6G-enabled networked intelligent systems, which establishes a nonlinear model between multi-modal sensing and channel characteristics, including large-scale and small-scale channel characteristics.
The rest of this paper is organized as follows. This paper first provides a concise review of existing channel modeling approaches, examining their core concepts, current applications, and inherent limitations, while identifying new modeling requirements for future networked intelligent systems. We then propose a novel MMICM framework, detailing its architecture and innovative features. Key enabling technologies and critical challenges, including data enhancement, multi-source data fusion, and large language model (LLM), are systematically analyzed. Finally, promising future research directions are discussed.

\section{Channel Modeling for Future Networked Intelligent Systems}
\begin{figure*}[t]
\centering
\includegraphics[width=1\textwidth]{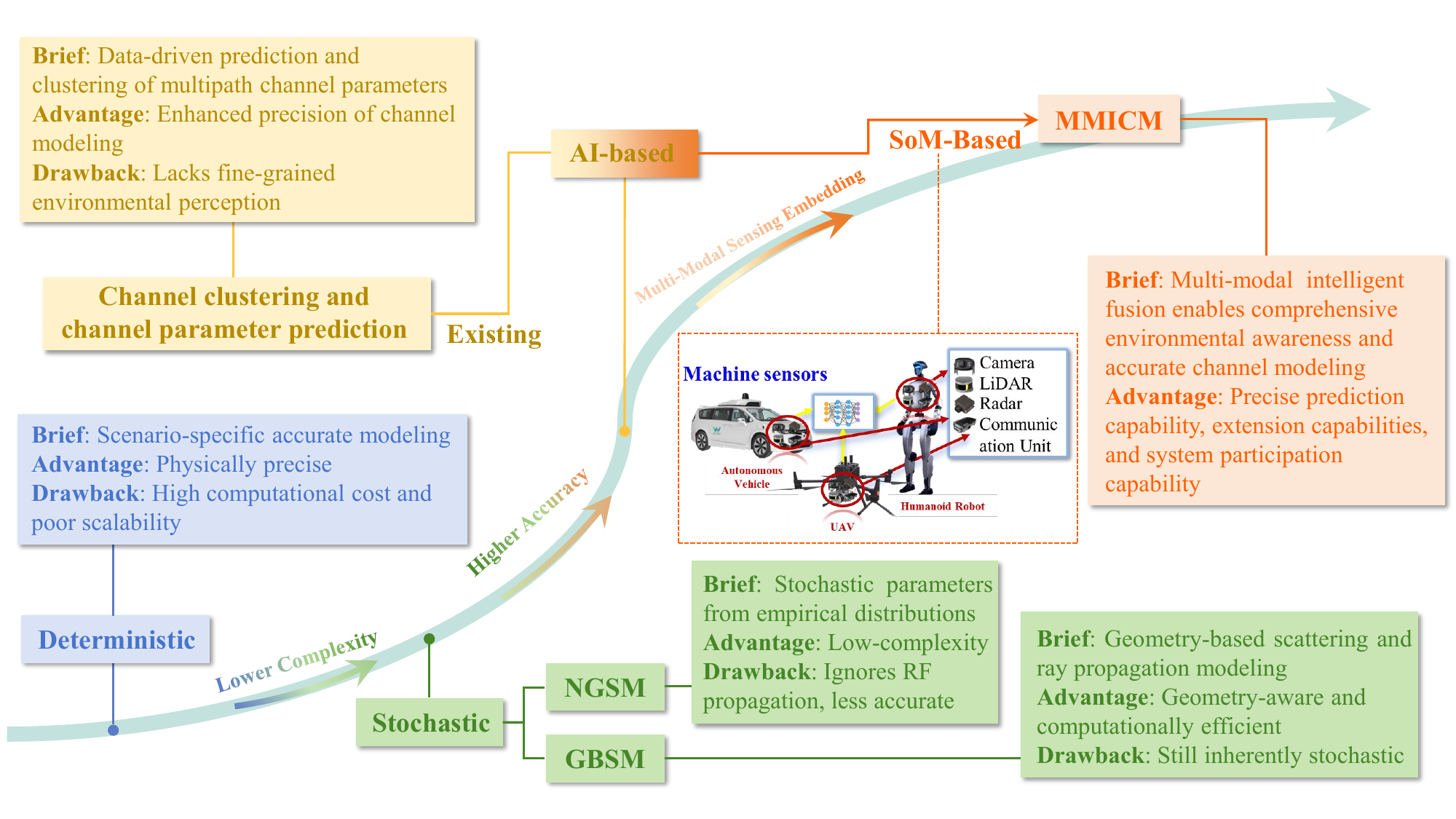}
\caption{The evolution of channel modeling methodologies.}
\label{t1} 
\end{figure*}

This section first reviews mainstream channel modeling methodologies, followed by a focused comparative analysis of representative AI-based channel modeling as shown in Fig.~\ref{t1}. Finally, we critically examine the limitations and challenges of existing methods in light of emerging requirements for 6G-enabled networked intelligent systems.

\subsection{Mainstream Channel Modeling Methods}
At present, the mainstream channel modeling methods includes deterministic channel modeling and stochastic channel modeling. 

Deterministic channel models offer site-specific characterizations by fully reconstructing electromagnetic wave propagation via measurements or ray-tracing, but suffer from high computational cost due to scenario-dependent precision.
For complexity reduction, stochastic channel models characterize wireless channels through statistical parameters, categorized as either non-geometric (NGSMs) or geometry-based stochastic models (GBSMs). NGSMs rely on empirical distributions (e.g., Rayleigh, Ricean, Nakagami-m) to model channel parameters but treat the channel as a statistical black box, neglecting the physical scattering processes in the environment.

To address this limitation, GBSMs incorporate predefined stochastic distributions of scattering clusters while explicitly modeling ray propagation in the environment. Based on geometric configurations of scattering propagation, GBSMs can be categorized into regular-shaped (RS-GBSMs) and irregular-shaped geometries (IS-GBSMs). 
RS-GBSMs enforce geometric regularity by placing scatterers on predefined 2D/3D shapes (e.g., semi-spheres, cylinders, ellipsoids) for theoretical analysis. In contrast, IS-GBSMs offer greater flexibility by allowing scatterer locations to follow statistical distributions rather than fixed shapes, making them well-suited for high-mobility scenarios.
Due to its favorable accuracy-complexity trade-off, IS-GBSM has been widely adopted in current 5G standardized channel models \cite{cx-wang1}, e.g., European COoperation in the field of Scientific and Technical research (COST) 2100 channel model, Third Generation Partnership Project (3GPP) TR38.901 channel model, Mobile and wireless communications Enablers for the Twenty-twenty Information Society (METIS) channel model, IMT-2020 channel model, and QUAsi-Deterministic RadIo channel GernerAtor (QuaDRiGa) channel model.

\subsection{AI-Based Channel Modeling Methods}
Regarding wireless communications evolution, channel modeling is expected being more and more accurate to advance in three key directions: (1) expanded scenario characterization encompassing outer space, aerial, maritime, and underwater environments; (2) enhanced support for diverse transceiver platforms including unmanned aerial vehicles (UAVs) and vehicular systems; and (3) extended frequency coverage spanning from sub-6GHz to terahertz (THz) bands. 
Furthermore, the emerging paradigm of AI-native communication systems further highlights the central role of massive high-fidelity data, since the scale and quality of training datasets fundamentally determine the performance ceiling.
Consequently, the application of AI technologies has been incorporated into channel modeling studies, revitalizing mainstream channel modeling approaches through machine learning techniques \cite{miy}.

Existing AI-based channel modeling approaches can be categorized into two main directions: channel clustering and channel parameter prediction.
In channel clustering, machine learning algorithms group multipath components (MPCs) with similar characteristics, enabling modeling of MPC evolution within propagation channels. This enhances the accuracy of non-stationarity and temporal consistency characterization. MPC clustering also serves as a key preprocessing step that reveals underlying data patterns and improves overall modeling precision.
In channel parameter prediction, machine learning is used for both large-scale fading (e.g., path loss) and small-scale fading estimation, including parameters such as arrival angles and time delays. Accurate prediction of these parameters significantly improves channel modeling precision.

\subsection{Challenges and Opportunities}

The emergence of 6G-enabled networked intelligent systems imposes significantly more stringent requirements and presents new challenges for channel modeling \cite{cx-wang2}. 
First, to support real-time collaboration and distributed decision-making in intelligent systems, next-generation channel modeling must integrate real-time processing with high-precision modeling to enable dynamic channel state analysis for time-critical operations.
Second, it should provide strong generalization across frequencies and scenarios, allowing efficient and low-overhead adaptation to both typical and extreme environments, thereby ensuring reliable connectivity.
Third, it needs to incorporate physical environment awareness, such as aided sensing and positioning, to enhance cognitive decision-making capabilities in intelligent systems.

The high computational complexity of deterministic channel modeling limits its ability to generalize across frequencies and scenarios in 6G-enabled intelligent systems. In contrast, the inherent randomness of stochastic modeling makes it unsuitable for real-time, high-precision channel analysis.
AI-based approaches that rely solely on RF data and panoramic maps lack detailed environmental perception, resulting in limited support for high-mobility scenarios and mission-critical communications requiring ultra-reliable, low-latency performance.

Fortunately, intelligent agents in intelligent systems (e.g., autonomous vehicles, UAVs, robots) inherently integrate multi-modal sensors and communication modules, serendipitously offering transformative opportunities for channel modeling to address the aforementioned challenges. Beyond wireless communication unit, these agents carry diverse sensors, including  millimeter wave (mmWave) radar, Light Detection And Ranging (LiDAR),  RGB imaging, and depth maps. The cross-modal sensing robustness mitigates mmWave’s limitations while their informational richness enhances environmental characterization.
Inspired by human synesthesia, we conceptualize multi-modal sensors and communication devices as an agent’s ‘sensory organs’ and artificial neural networks as its ‘brain’, proposing Synesthesia of Machines (SoM) \cite{som}. SoM intelligently fuses communications with multi-modal perception, unlocking symbiotic synergies between modalities.
Leveraging this paradigm, we investigate the physics-to-electromagnetics mapping mechanism by exploiting multi-modal sensing’s synergistic enhancement of channel modeling, and proposed a novel MMICM framework for 6G-enabled networked intelligent systems.


\section{Novel Multi-Modal Intelligent Channel Modeling Framework and Key Technologies}
Leveraging the multi-modal sensors equipped on the intelligent agents, the proposed MMICM framework explores the mapping relationship between physical environment and electromagnetic channel via SoM, and further achieves precise prediction capability, extension capabilities, and system participation capability for 6G-enabled networked intelligent systems. As a data-driven approach, this framework critically depends on massive volumes of high-quality data, with its modeling accuracy being fundamentally constrained by dataset integrity.
\subsection{Dataset Construction}
\begin{figure}[t]
\centering
\includegraphics[width=0.49\textwidth]{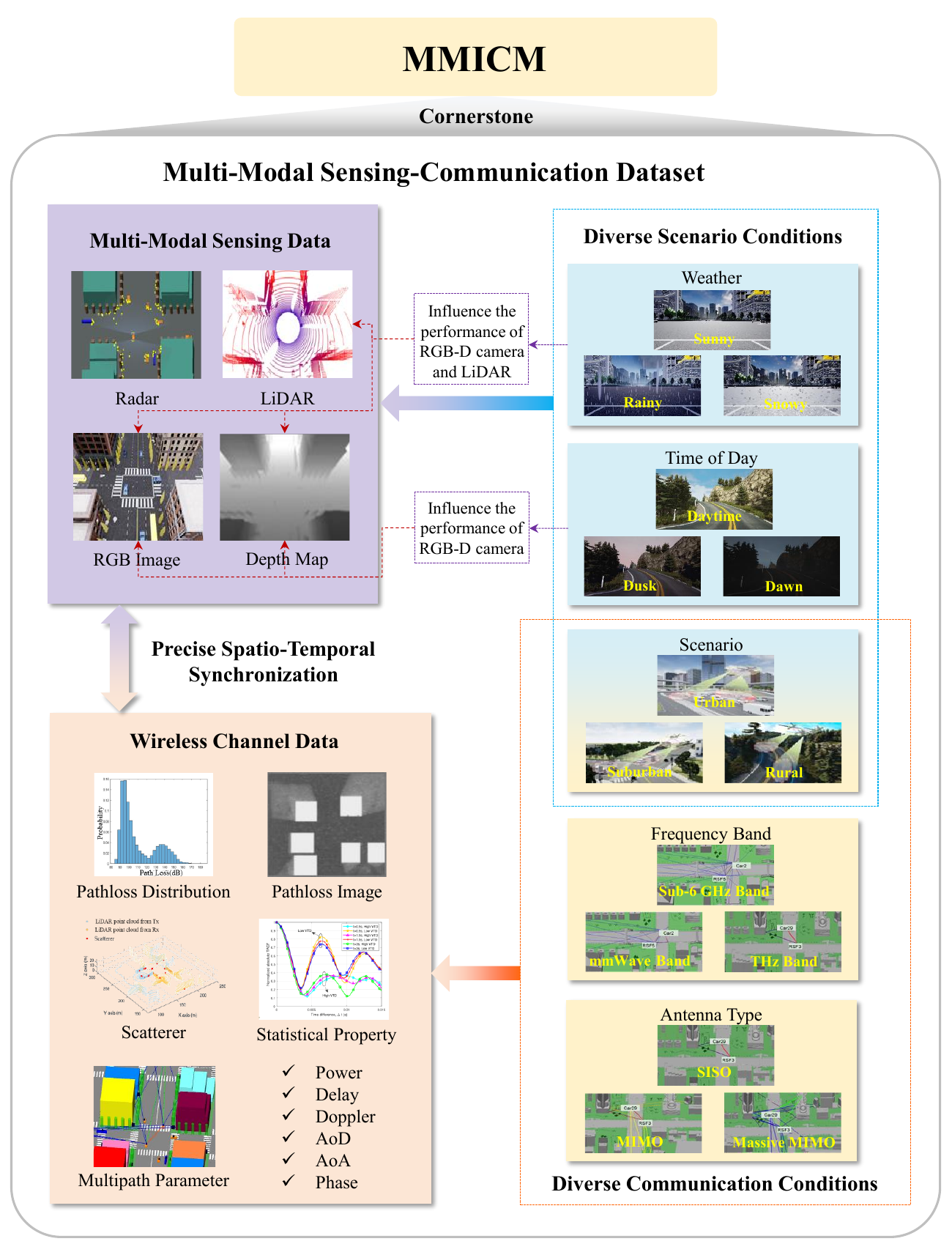}
\caption{The multi-modal sensing-communication datasets.}
\label{t4} 
\end{figure}

As the cornerstone of MMICM, the comprehensive multi-modal sensing-communication datasets are composed of multi-modal sensing data (e.g., mmWave radar point clouds, LiDAR point clouds, RGB images,
and depth maps) and wilreless communications channel data (e.g., large scale channel parameters and small scale channel parameters), as illustrated in Fig.~\ref{t4}.
The construction of comprehensive multi-modal sensing-communication dataset necessitates a sophisticated hardware ensemble comprising both wireless transceivers and multi-modal sensors (e.g., mmWave radar, RGB-D cameras, LiDAR). The foremost technical priorities of multi-modal sensing-communication datasets lies in achieving precise spatio-temporal synchronization across these heterogeneous devices, requiring sub-millisecond timing alignment and centimeter-level spatial registration \cite{dataset}. 
Another critical technical focus of multi-modal sensing-communication datasets is the development of richly diverse datasets that comprehensively span 6G-enabled networked intelligent environments. 
First, it necessitates data collection across varied geographic scenarios of 6G-enabled networked intelligent systems including urban, suburban, rural, indoor, and so on, while simultaneously incorporating key 6G technologies, for example massive multiple-input multiple-output (MIMO) deployments. 
Second, the datasets must further capture diverse frequency-dependent propagation characteristics through inclusive coverage of sub-6~GHz, millimeter-wave (24-100~GHz) and terahertz ($\textgreater$100~GHz) bands to account for their distinct channel behaviors. 
Third, the high-dynamic mobility patterns of 6G-enabled networked intelligent systems require special attention, particularly UAV operations across altitude strata (10-500~m) and automated guided vehicle (AGV) movements under varying vehicle traffic densities. Finally, to ensure robust multi-modal sensor performance, datasets must systematically encompass diurnal lighting variations (dawn/noon/dusk) and adverse weather conditions (rain/snow) while addressing each sensor modality's unique environmental sensitivities.

\subsection{Modeling Structure}
\begin{figure*}[t]
\centering
\includegraphics[width=1\textwidth]{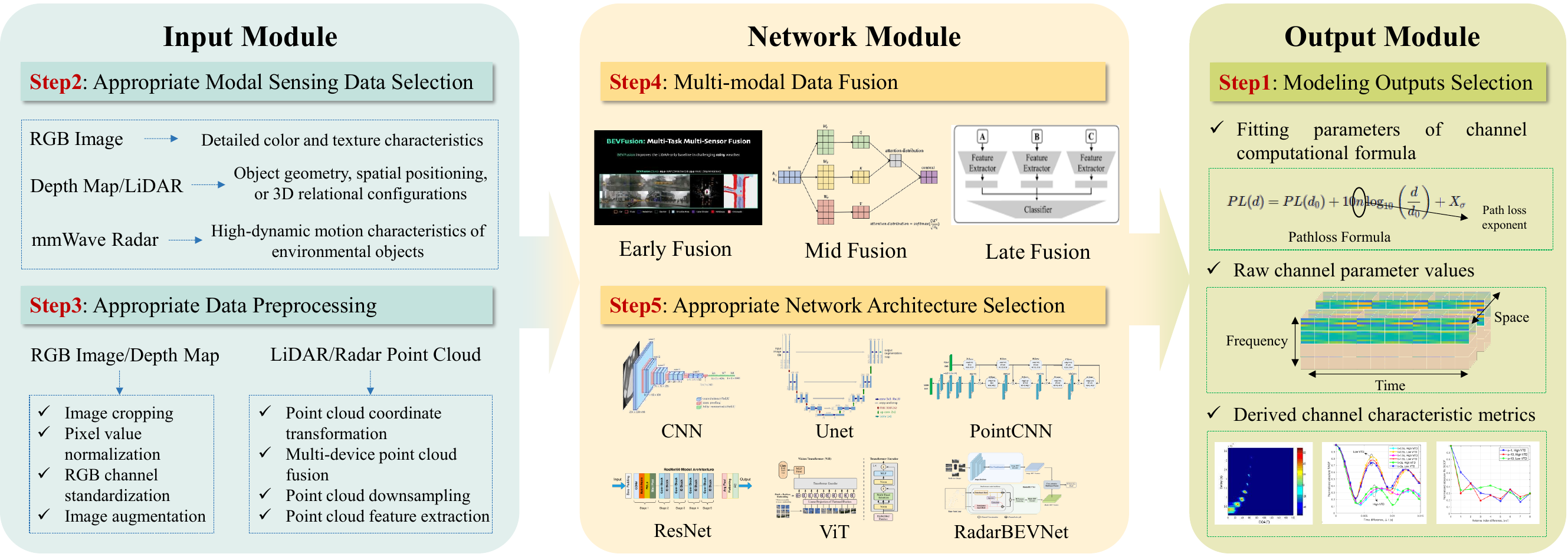}
\caption{The overall structure of proposed MMICM framework.}
\label{t2} 
\end{figure*}

To explore the mapping relationship between physical environment and electromagnetic channel, MMICM framework comprises three fundamental components: input module, output module, and network model architecture. Fig.~\ref{t2} illustrates the overall structure of proposed MMICM framework. Red texts mark 5 key steps. 

\subsubsection{Output Module}
The output module plays a pivotal role in MMICM framework, as it fundamentally defines the model's operational objectives and functional capabilities.

\textbf{Step 1:} 
Depending on the specific modeling objectives and intended downstream tasks, appropriate modeling outputs should be selected, encompassing both large-scale and small-scale channel information. 

With selection contingent upon application requirements and computational constraints, the specific output formats include fitting parameters of channel computational formula (e.g., path loss exponents, shadowing variances), raw channel parameter values (e.g., CSI matrices), or derived channel characteristic metrics (e.g., delay spread, K-factor).
Path loss modeling by MMICM has evolved from the early fitting parameters prediction in channel formulas to the current path loss image prediction where each pixel in the path loss image represents a point-to-point path loss value. It means that, by leveraging multi-modal sensing data which provides more specific and detailed environmental information, MMICM can simultaneously predict hundreds of point-to-point path loss values within a localized area at a resolution of few meters. It is an unprecedented capability in terms of both functionality and precision, which can support closed-loop optimization of both communication links and UAV/AGV trajectories in 6G-enabled networked intelligent systems.
Small-scale channel information (e.g., multipath components, Doppler shifts)
For small-scale channel information, the multipath propagation typically involves numerous paths, each characterized by parameters such as power, phase, angle of departure (AoD), angle of arrival (AoA), and Doppler shift. Given this complexity, MMICM often employs statistical properties as the output.

\subsubsection{Input Module}
The input module is designed for core environment information extraction, serving as the foundational data feature extraction component of the MMICM framework. 

\textbf{Step 2:} 
Select appropriate modal sensing data from multi-modal sensing information as input of MMICM.

First, when the modeling outputs have strong correlation with environmental objects and their material properties, RGB images emerges as the preferred input due to its inherent capacity to capture detailed color and texture characteristics. The photometric richness of RGB data provides high semantic density that is particularly well-suited for object-centric analysis and material classification tasks, offering superior performance in most conventional visual recognition scenarios compared to non-visual sensing modalities. 
Second, when modeling outputs have strong correlation with object geometry, spatial positioning, or 3D relational configurations, depth maps or LiDAR point clouds constitute a more suitable input due to their inherent capacity for 3D spatial information. 
Third, when modeling outputs necessitate capturing high-dynamic motion characteristics of environmental objects, mmWave radar emerges as the optimal modality due to its unparalleled capability for real-time tracking and dynamic state analysis of target entities.
Leveraging Doppler shift, mmWave radar enables accurate velocity estimation and trajectory inference, which are essential for modeling channel dynamics such as Doppler spread and path evolution. Compared to visual modalities, it maintains robust performance under low visibility conditions like darkness, fog, and rain. Its strong penetration and narrow beamwidth further support high-resolution spatial perception, even for partially occluded targets.

\textbf{Step 3:} 
Perform appropriate data preprocessing on the selected modal data to ensure optimal feature extraction and downstream analysis.

For RGB images and depth maps, essential data preprocessing operations include image cropping, pixel value normalization, RGB channel standardization, and image augmentation, which are critical for ensuring data quality and model effectiveness. Image cropping serves to standardize dimensional attributes across input images, maintaining consistent aspect ratios critical for neural network learning. Pixel value normalization rescales input data to either the $[0,1]$ or $[-1,1]$ range, serving dual purposes of accelerating model convergence and mitigating gradient instability issues, such as explosion and vanishing gradients. RGB channel standardization is a preprocessing technique that normalizes the red (R), green (G), and blue (B) channels of an image, ensuring numerical stability and reproducible model performance.
For LiDAR point clouds and radar point clouds, point cloud coordinate transformation, multi-device point cloud fusion, point cloud downsampling, and point cloud feature extraction represent essential preprocessing operations. Commonly employed point cloud downsampling methods include voxel grid filtering, random sampling, uniform sampling, and k-nearest neighbors (KNN)-based sampling, each offering distinct computational-performance tradeoffs for 3D point cloud data simplification. 

\subsubsection{Network Model}
The network model should be co-designed with task requirements and data characteristics. A well-constructed network model is crucial for MMICM, as it enables more effective extraction of the underlying mapping relationships between physical environment and electromagnetic propagation.

\textbf{Step 4:} 
Proceed with multi-modal data fusion (if applicable) to integrate complementary information across heterogeneous sensory inputs.

In MMICM, multi-modal fusion strategies can be broadly classified into early, mid, and late fusion, depending on the stage at which data from different modalities are integrated. Early fusion operates at the input level by spatially aligning raw sensor data, such as projecting RGB images and LiDAR point clouds onto a common Bird's Eye View (BEV) or image plane. This alignment enables direct concatenation and facilitates low-level feature aggregation. Mid fusion is the most prevalent approach, emphasizing cross-modal interaction at the feature representation stage. Common techniques include channel attention and cross-attention mechanisms. Channel attention adaptively assigns weights to features from different modalities based on their relative importance, thereby enhancing salient information while suppressing less informative signals. Cross-attention enables complementary feature learning by allowing one modality to query relevant features from another. In these mid-fusion architectures, self-attention modules within Transformer frameworks are typically employed to model dependencies within or across fused representations. Finally, late fusion combines the outputs or high-level features from individual modality-specific branches after independent processing, making it well-suited for modular design or task-decoupled scenarios.

\textbf{Step 5:} 
Select appropriate network architectures.

Convolutional neural networks (CNN), ResNet, UNet, and Vision Transformer (ViT) are widely used in MMICM for visual sensing data such as RGB images and depth maps.
CNNs use convolutional layers to extract spatially localized features, making them suitable for low-level feature extraction in RGB images. Their simple implementation and efficiency make them ideal for resource-constrained tasks.
ResNet is effective for classification, detection, and segmentation, with skip connections that enable deep network training while avoiding gradient issues. 
UNet, with its encoder-decoder structure and skip connections, is designed for pixel-level segmentation, excelling in tasks like medical imaging and scene parsing by accurately recovering spatial structure.
ViT, based on the Transformer architecture, captures global context and performs strongly on large-scale image tasks. It outperforms CNNs on large datasets and long-range dependency modeling, making it ideal for global pattern recognition.

PointNet++, PointCNN, and RadarBEVNet are commonly used in MMICM for processing point cloud data such as mmWave radar and LiDAR.
PointNet++ is suited for 3D object recognition, segmentation, and classification, especially with complex and sparse data. It captures local features at multiple scales and adapts well to large-scale, unordered point clouds with complex geometries.
PointCNN, also used for classification, segmentation, and reconstruction, applies convolution directly to point clouds, better preserving structural relationships. Compared to PointNet++, it offers higher accuracy in object detection and spatial modeling.
RadarBEVNet is tailored for RGB-radar fusion in BEV-based detection. Its dual-stream radar backbone and RCS-aware encoder are optimized for sparse radar signals, while deformable cross-attention ensures accurate alignment and efficient fusion with camera data.

Recent advances in large-scale visual pretraining and cross-modal modeling have introduced more powerful neural architectures into MMICM.
For visual modalities, Transformer-based models such as Swin Transformer, SegFormer, and Masked Autoencoder (MAE) capture long-range dependencies while preserving hierarchical structure, enabling fine-grained spatial understanding in complex scenes. Pretrained models like Contrastive Language–Image Pretraining (CLIP) and DINO further improve semantic alignment across modalities.
For point cloud data, newer Transformer-based encoders like Point-BERT, Point-MAE, and PointNeXt use patch-level tokenization and self-supervised learning to capture detailed geometric and relational features. Lightweight radar-specific encoders such as RadarFormer and RadarCLIP emphasize efficient representation and semantic consistency with visual data.
These architectures provide a solid foundation for robust multi-modal fusion and cross-modal representation in MMICM.

\begin{figure}[t]
\centering
\includegraphics[width=0.5\textwidth]{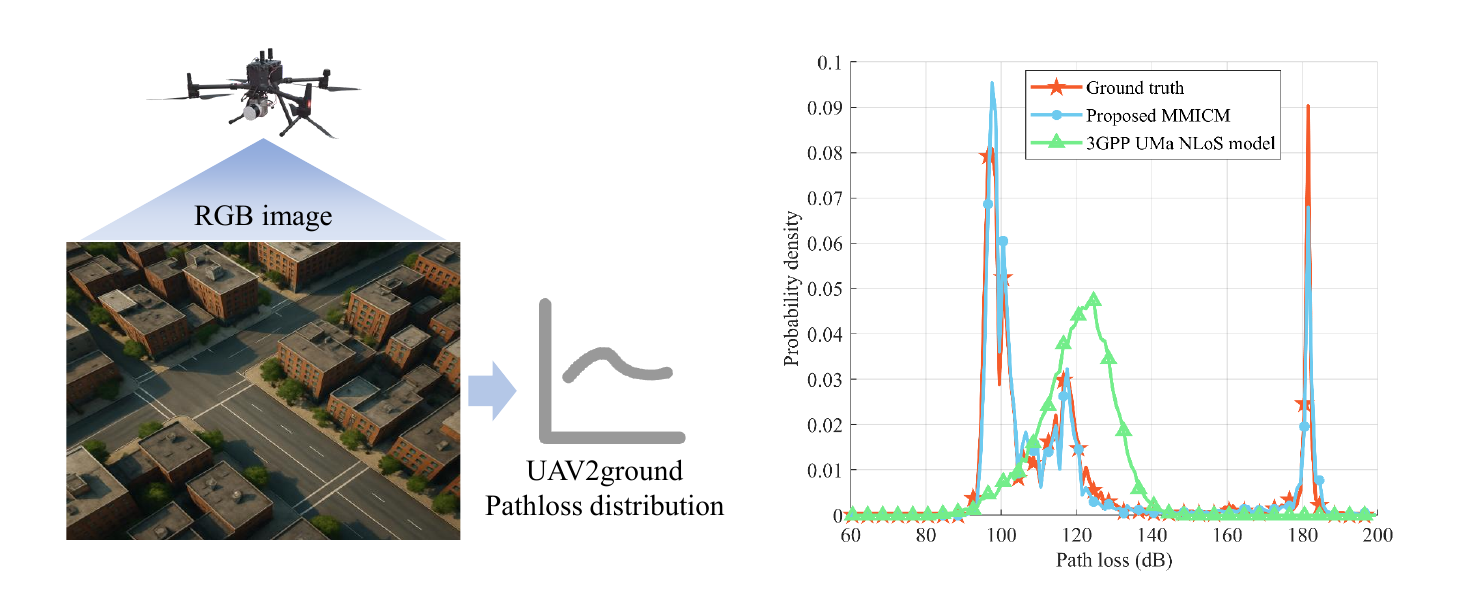}
\caption{The RGB image captured by the UAV serves as the input to infer the U2G pathloss distribution (left). The probability density distributions of the pathloss generated by the ground truth, the proposed MMICM model, and the 3GPP UMa NLoS model are compared under the same scenario (right).}
\label{pl} 
\end{figure}

\begin{figure*}[t]
\centering
\includegraphics[width=1\textwidth]{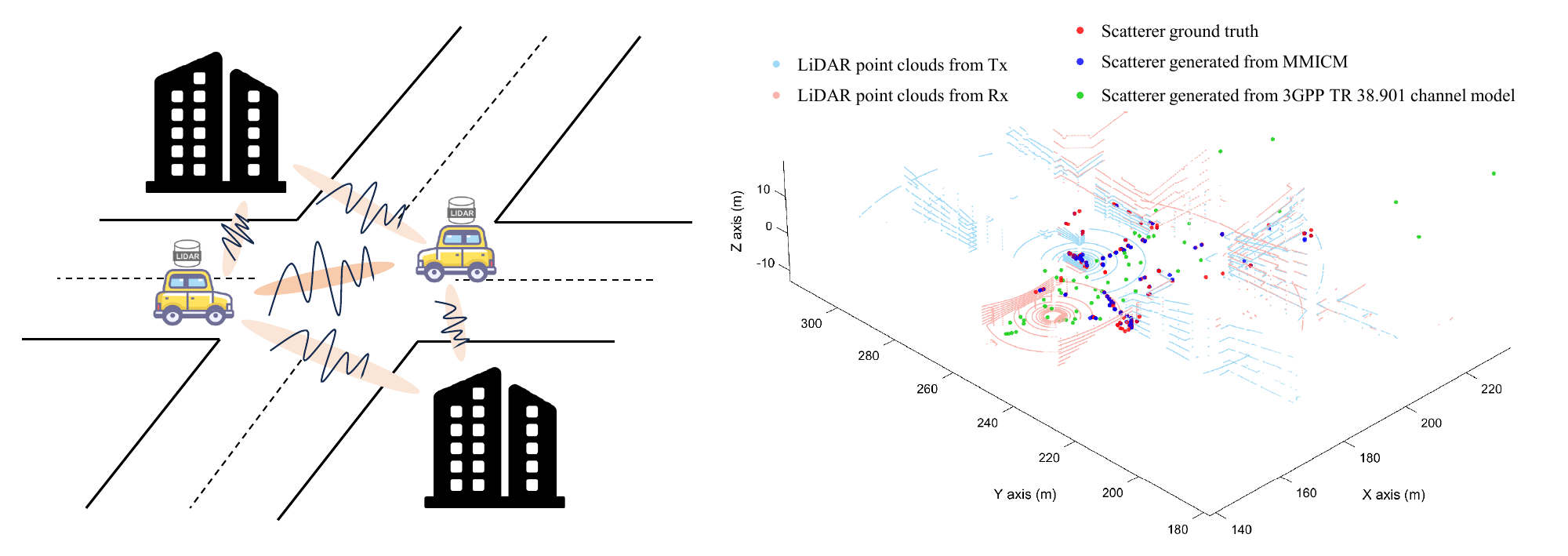}
\caption{In the V2V scenario, LiDAR sensors mounted on transceivers capture the surrounding environment as input for scatterer generation (left). The 3D distributions of scatterers from the ground truth, the proposed MMICM, and the 3GPP TR 38.901 model are compared (right).}
\label{scatterer} 
\end{figure*}

To validate the advantages of proposed MMICM, we present two modeling examples that highlight its effectiveness in modeling scene-specific wireless propagation, including large-scale modeling and small-scale modeling.
Fig.~\ref{pl} shows the UAV-to-ground (U2G) pathloss distribution in an urban scenario, where an aerial RGB image captured by the UAV is utilized to infer the corresponding pathloss distribution. The result is compared with the ground truth and the conventional 3GPP UMa NLoS model. MMICM produces a pathloss distribution that closely matches the ground truth, particularly in both low- and high-loss regions, while the 3GPP model shows significant deviation. This improvement stems from MMICM’s ability to extract visual semantics and structural features from the real environment, whereas the 3GPP model relies on statistical assumptions without considering specific environmental geometry, leading to limited accuracy.
Fig.~\ref{scatterer} illustrates the scatterer generation in a V2V urban crossroad scenario, where both the transmitter and receiver vehicles are equipped with LiDAR sensors to perceive the surrounding environment. According to the mapping between physical environment and electromagnetic space which is explored by MMICM, the LiDAR point clouds are utilized to generate the spatial distribution of electromagnetic scatterers. Compared to the ground truth, MMICM accurately reconstructs the 3D positions of scatterers with better spatial consistency and density, while the scatterers generated by the 3GPP TR 38.901 channel model are sparse and misaligned with the actual scene layout. This is because MMICM leverages detailed geometric cues from the LiDAR point clouds, capturing the true propagation-related structures, whereas the conventional model adopts a stochastic approach that overlooks scene-specific features.
Overall, these results highlight MMICM’s strength in embedding physical environment knowledge into channel modeling. By utilizing rich multi-modal sensing inputs such as RGB images and LiDAR data, MMICM significantly improves the accuracy and environmental awareness of both large-scale and small-scale channel characteristics.


\section{Highly System-Engaged Applications}
\begin{figure*}[t]
\centering
\includegraphics[width=1\textwidth]{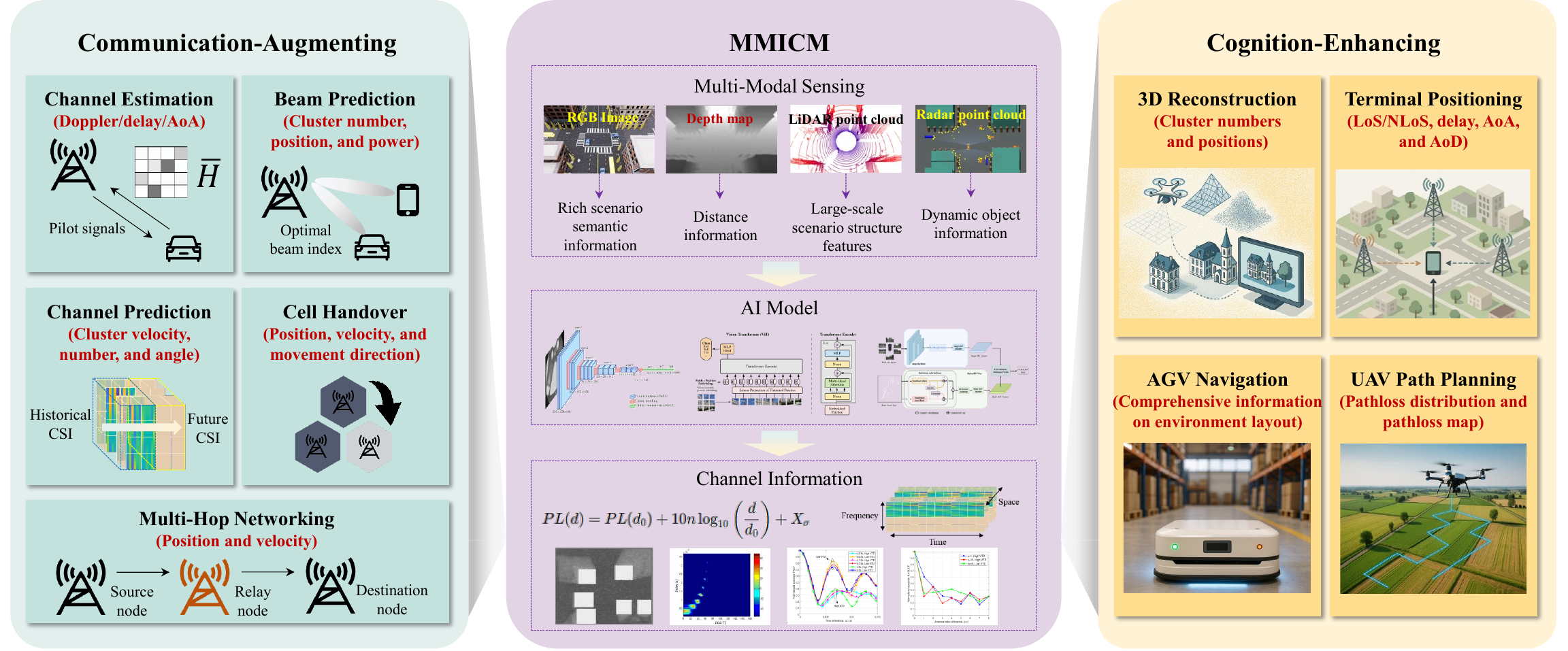}
\caption{Highly system-engaged applications of proposed MMICM framework.}
\label{t3} 
\end{figure*}

Leveraging multi-modal sensing that better characterizes physical environmental details, the proposed intelligent modeling framework not only achieves real-time precise prediction of high-dynamic channel states and cross-frequency/scenario generalization, but also introduces a groundbreaking capability, i.e., highly system engagement. 
In this section, the highly system-engaged applications of MMICM are discussed through two distinct aspects: communication-augmenting applications and cognition-enhancing applications, as illustrated in Fig.~\ref{t3}. It is worth noting that our framework also holds the potential to generate massive high-quality data, which further boosts the performance of these applications.

\subsection{Communication-Augmenting Applications}
Benefiting from in-depth investigations into the mapping relationships between multi-modal sensing and channel information, MMICM acquires significantly more actionable information compared to conventional channel modeling approaches, e.g., cluster characteristics (number, power, position, etc.), multipath component (MPC)-related velocity profiles, position, velocity, and movement of user equipments (UEs) and dynamic objects. This capability enables MMICM to actively support critical communication-augmenting applications, including channel estimation, beam prediction, channel prediction, cell handover, and multi-hop networking.

For channel estimation, leveraging multi-modal information from MMICM enables accurate estimation of cluster numbers and positions. This allows more precise initialization of the Orthogonal Matching Pursuit (OMP) algorithm and reduces the iteration count required for high-accuracy channel state information (CSI) recovery \cite{channelestimation}.
For beam prediction, environmental semantics extracted from street camera data are transmitted to the base station (BS) for initial beam index prediction. Integrating MMICM's multi-modal capabilities, which provide cluster-level information (number, position, and power), enriches the semantic representation and improves prediction accuracy.
Channel prediction forecasts future channel behavior based on current and historical characteristics. MMICM enables extraction of geometric details, such as MPC count and angles-of-arrival (AoAs), which are critical for accuracy, especially in dynamic scenarios with rapidly changing propagation conditions.
Cell handover ensures uninterrupted call or data transfer between cells \cite{handover}. MMICM allows accurate prediction of UE parameters (e.g., position, velocity, trajectory) through multi-modal sensing, enhancing situational awareness and enabling more efficient, low-latency handover decisions in dynamic environments.
For multi-hop networking, MMICM uses multi-modal data (RGB images, depth maps, LiDAR, and mmWave radar) to precisely track dynamic object positions and velocities. This supports two key functions: velocity estimation enables Doppler offset correction, improving link reliability; position tracking allows dynamic relay node organization, optimizing end-to-end routing.
\subsection{Cognition-Enhancing Applications}
Leveraging the mapping between multi-modal sensing and channel information, MMICM demonstrates superior environmental awareness capabilities compared to conventional channel modeling methods. Specifically, it enables real-time path loss predition, 
Line-of-sight (LoS) and non-line-of-sight (NLoS) path identification, comprehensive cluster characterization (number, power distribution, spatial configuration), and radio wave-object interaction analysis. 
These advanced capabilities of MMICM empower it to actively facilitate several cognition-augmenting applications, including 3D reconstruction, mobile terminal positioning, AGV path planning, and UAV path planning.

For 3D reconstruction, MMICM-based cluster analysis (e.g., number and spatial distribution) extracts the topological layout of objects and environmental features, providing valuable geometric priors to improve reconstruction accuracy.
For mobile terminal positioning, MMICM’s multi-modal sensing enables accurate identification of LoS/NLoS paths by analyzing cluster positioning data, mitigating excess delay errors inherent in conventional RF-only methods and significantly improving positioning precision.
Conventional AGV navigation systems relying only on onboard sensors are limited in perceiving obstacle geometries and environmental layouts \cite{AGVPath}. MMICM overcomes this by integrating multi-modal sensing with communication channel information. Through analysis of radio wave-object interactions, it achieves comprehensive spatial awareness, including detailed obstacle profiling and environmental mapping, supporting optimized trajectory generation, dynamic obstacle avoidance, and improved navigation efficiency.
For UAV path planning, the key challenge is jointly optimizing communication performance and energy efficiency \cite{UAVPath}. While conventional models are computationally expensive due to iterative optimization, MMICM leverages real-time multi-modal sensing to dynamically estimate path loss and spatial features, enabling closed-loop optimization of both communication links and trajectories with lower computational overhead.

\section{Future Research Directions}
The open challenges and future research directions for MMICM are systematically discussed through three critical dimensions, including data acquisition and processing, network architecture design, and practical application scenarios.

\textbf{Data Acquisition and Processing:} 
While data-driven channel modeling already requires large datasets across multiple scenarios and frequency bands, MMICM further demands precisely aligned multi-modal sensing data, greatly increasing the complexity of data acquisition. Artificial intelligence generated content (AIGC), powered by generative AI (GAI), offers strong data synthesis capabilities \cite{msheng}. Using diffusion models with zero-shot transfer learning to generate large-scale multi-modal datasets is a key research direction.
Producing flexible and comprehensive output representations remains a major challenge. Due to the many propagation paths and the high-dimensional parameter space of each component, current MMICM outputs are limited to partial estimations and statistical descriptions. Achieving unified modeling of all propagation parameters is both highly demanding and urgently needed.

\textbf{Network Architecture Design and Transferring:} 
While stochastic channel modeling offers broad applicability and low complexity, it struggles with real-time, site-specific channel prediction. In contrast, MMICM enables channel prediction through multi-modal sensing, but its generalization remains limited by data-driven constraints. A promising direction is to develop hybrid architectures that combine neural networks with statistical models, integrating the strengths of both approaches while improving interpretability and providing actionable insights for future development.
Current methods (e.g., CLIP) still show semantic gaps in cross-domain generalization when applied to zero/few-shot scenarios and frequency adaptation, highlighting the need for novel generalization strategies tailored to multi-modal sensing-communication fusion in MMICM.

\textbf{Practical Application Scenarios:} 
A Digital Twin (DT) serves as a virtual counterpart to physical entities and systems \cite{DT}. By leveraging real-time DT technology, 6G-enabled intelligent systems can achieve dynamic decision-making with near-zero latency, enabling closed-loop interaction between physical and digital domains. Similarly, MMICM can map the electromagnetic space of communication channels to the physical space of multi-modal sensing. Fully utilizing MMICM’s strengths to support DT development is a promising research direction.
The emergence of 6G-enabled intelligent systems signals a shift from “Internet intelligence” to “Embodied intelligence” ecosystems \cite{embody}. This transition demands ultra-precise environmental perception, ultra-reliable low-latency communication, and robust operation in dynamic scenarios involving concurrent mobility and interaction among multiple agents. These requirements intensify the challenges in exploring MMICM’s cross-domain mapping.

\section{Conclusions}
The development of 6G-enabled networked intelligent systems demands accurate and real-time channel modeling to support emerging requirements, such as precise prediction capability, extension capabilities at diverse scenarios and frequency bands, and system participation capability. 
Leveraging the multi-modal sensors equipped on the intelligent agents, the proposed MMICM framework overcomes these challenges by explore the mapping relationship between physical environment and electromagnetic channel through AI-driven techniques. 
Benefiting from its capability to fully exploit the mapping relationships between multi-modal sensing and communications via SoM, MMICM can support both communication-augmenting applications (channel estimation, beam prediction, channel prediction, cell handover, and multi-hop networking) and cognition-enhancing applications (3D reconstruction, mobile terminal positioning, AGV path planning, and UAV path planning).
Moreover, MMICM inherently holds the advantage of generating massive high-quality data, which is crucial for AI-native communication systems whose performance upper bound is fundamentally determined by data scale and fidelity.
In the future, the research on MMICM still faces critical challenges across three key domains, such as data acquisition and processing, network architecture design, and practical application scenarios.

\vfill


\begin{thebibliography}{29}
\bibitem{NIS}
R. Liu \emph{et al.}, ``6G enabled advanced transportation systems," \emph{IEEE Trans. Intelligent Transportation Systems}, vol. 25, no. 9, pp. 10564--10580, Sept. 2024.

\bibitem{6}
A. Molisch, \emph{Wireless Communications}. UK: John Wiley Sons, 2011.

\bibitem{survey-AI}
A. Seretis and C. D. Sarris, ``An overview of machine learning techniques for radiowave propagation modeling," \emph{IEEE Trans. Antennas Propag.}, vol. 70, no. 6, pp. 3970–-3985, June 2022.








\bibitem{cx-wang1}
C.-X. Wang \emph{et al.}, ``A survey of 5G channel measurements and models,"  \emph{IEEE Commun. Surveys \& Tutorials}, vol. 20, no. 4, pp. 3142–-3168, Fourth--quarter 2018.

\bibitem{miy}
M. Yang \emph{et al.}, ``AI-enabled data-driven channel modeling for future communications,"  \emph{ IEEE Commun. Mag.}, vol. 62, no. 4, pp. 112--118, Apr. 2024.



\bibitem{cx-wang2}
L. Bai \emph{et al.}, ``Multi-modal intelligent channel modeling: A new modeling paradigm via synesthesia of machines,"  \emph{IEEE Commun. Surveys \& Tutorials}, 2025. DOI: 10.1109/COMST.2025.3558046


\bibitem{som}
X. Cheng \emph{et al.}, ``Intelligent multi-modal sensing-communication integration: Synesthesia of Machines," \emph{IEEE Commun. Surveys \& Tutorials}, vol. 26, no. 1, pp. 258--301, first---quarter 2024.


\bibitem{dataset}
X. Cheng \emph{et al.}, ``SynthSoM: A synthetic intelligent multi-modal sensing-communication dataset for Synesthesia of Machines (SoM)," \emph{Sci. Data}, 2025. [Online]. Available: https://arxiv.org/abs/2501.07459.


\bibitem{channelestimation}
S. Jiang and A. Alkhateeb, ``Sensing aided OTFS massive MIMO systems\: Compressive channel estimation," in \emph{Proc. IEEE Int. Conf. Commun. Workshops (ICC Workshops)}, Rome, Italy, Jun. 2023, pp. 794–-799.

\bibitem{handover}
M. Zaher, E. Bjornson, and M. Petrova, ``Soft handover procedures in mmWave cell-free massive MIMO networks,"  \emph{IEEE Trans. Wireless Commun.}, vol. 23, no. 6, pp. 6124–-6138, Jun. 2024.

\bibitem{AGVPath}
J. Xin, \emph{et al.}, ``Model predictive path planning of AGVs\: Mixed logical dynamical formulation and distributed coordination,"  \emph{ IEEE Trans. Intell. Transp. Syst.}, vol. 24, no. 7, pp. 6943-–6954, Jul. 2023.

\bibitem{UAVPath}
Y. Dong, C. He, Z. Wang, and L. Zhang, ``Radio map assisted path planning for UAV anti-jamming communications,"  \emph{  IEEE Signal Process Lett.}, vol. 29, pp. 607–-611, Feb. 2022.



\bibitem{msheng}
H. Wang, H. Li, M. Sheng, and J. Li, ``Collaborative fine-tuning of mobile AIGC models with wireless channel conditions," \emph{IEEE Wireless Commun.}, vol. 31, no. 4, pp. 32--38, Aug. 2024.

\bibitem{DT}
A. Alkhateeb, S. Jiang, and G. Charan, ``Real-time digital twins\: Vision and research directions for 6G and beyond," \emph{ IEEE Commun. Mag.}, vol.61, no. 11, pp. 128-–134, Nov. 2023.

\bibitem{embody}
A. Gupta, S. Savarese, S. Ganguli, and F. Li, ``Embodied intelligence via learning and evolution," \emph{ Nat. Commun.}, vol. 12, pp. 5721–-5732, Oct. 2021.

\end{thebibliography}
\end{document}